\documentstyle[aps,prb,twocolumn,epsf,floats]{revtex}
\draft
\begin{document}
\wideabs{
\title{Electron Correlations 
       in Partially Filled Lowest and Excited Landau Levels}
\author{
   Arkadiusz W\'ojs}
\address{
   Department of Physics, 
   University of Tennessee, Knoxville, Tennessee 37996 \\
   Institute of Physics, 
   Wroclaw University of Technology, Wroclaw 50-370, Poland}
\maketitle
\begin{abstract}
   The electron correlations near the half-filling of the lowest 
   and excited Landau levels (LL's) are studied using numerical 
   diagonalization.
   It is shown that in the low lying states electrons avoid pair 
   states with relative angular momenta ${\cal R}$ corresponding 
   to positive anharmonicity of the interaction pseudopotential 
   $V({\cal R})$.
   In the lowest LL, the super-harmonic behavior of $V({\cal R})$ 
   causes Laughlin correlations (avoiding pairs with ${\cal R}=1$)
   and the Laughlin--Jain series of incompressible ground states.
   In the first excited LL, $V({\cal R})$ is harmonic at short 
   range and a different series of incompressible states results.
   Similar correlations occur in the paired Moore--Read $\nu={5\over2}$ 
   state and in the $\nu={7\over3}$ and ${8\over3}$ states, all having 
   small total parentage from ${\cal R}=1$ and 3 and large parentage 
   from ${\cal R}=5$.
   The $\nu={7\over3}$ and ${8\over3}$ states are different from Laughlin 
   $\nu={1\over3}$ and ${2\over3}$ states and, in finite systems, occur 
   at a different LL degeneracy (flux).
   The series of Laughlin correlated states of electron pairs at 
   $\nu=2+2/(q_2+2)={8\over3}$, ${5\over2}$, ${12\over5}$, and ${7\over3}$ 
   is proposed, although only in the $\nu={5\over2}$ state pairing has 
   been confirmed numerically.
   In the second excited LL, $V({\cal R})$ is sub-harmonic at short 
   range and (near the half-filling) the electrons group into spatially 
   separated larger $\nu=1$ droplets to minimize the number of strongly 
   repulsive pair states at ${\cal R}=3$ and 5.
\end{abstract}
}

\section{Introduction}
%%%%%%%%%%%%%%%%%%%%%%%%%%%%%%%%%%%%%%%%%%%%%%%%%%%%%%%%%%%%%%%%%%%%%%%%
When a pure two-dimensional electron gas (2DEG) in a high magnetic field 
fills a fraction $\nu$ of a degenerate Landau level (LL), the nature of
the ground state (GS) and low lying excitations are completely determined 
by their (Coulomb) interaction.
The correlations induced by this interaction can be probed in transport 
or optical measurements, and, for example, the occurrence of non-degenerate 
incompressible liquid-like GS's\cite{laughlin-fqhe} at certain values of 
$\nu$ is responsible for the fractional quantum Hall (FQH) effect.
\cite{tsui,prange,stormer}
In the lowest ($n=0$) LL, the FQH effect is observed at various filling 
factors $\nu={1\over3}$, ${2\over3}$, ${2\over5}$ etc., all being simple 
odd-denominator fractions.
The origin of these fractions lies in the special form of (Laughlin) 
correlations\cite{laughlin-fqhe} which result from the short-range 
character of the Coulomb interaction pseudopotential\cite{haldane-pseudo,%
wojs-parentage,quinn} in the lowest LL.
The explanation of all the observed fractions involves identification 
of Laughlin incompressible GS's at $\nu=(2p+1)^{-1}$, where $p$ is 
an integer, and their elementary (quasiparticle) excitations,
\cite{laughlin-fqhe} and the observation that at certain fillings 
$\nu_{\rm QP}$ the quasiparticles form Laughlin incompressible 
GS's of their own.
\cite{haldane-hierarchy,laughlin-hierarchy,halperin-hierarchy}
This (Haldane's) hierarchy construction predicts no incompressible
GS's at even-denominator fractions, in perfect agreement with the 
experiments in the lowest LL.
Because of its equivalence\cite{sitko-hierarchy,wojs-hierarchy} to 
Haldane's hierarchy picture, Jain's non-interacting composite fermion 
(CF) model\cite{jain-cf,lopez,jain-hierarchy,halperin-lee-read} 
also predicts FQH states at the same fractions.

Quite surprisingly, the FQH effect at an even-denominator fraction 
has been discovered\cite{willet,eisenstein-review,eisenstein-tilt1,%
eisenstein-tilt2,eisenstein-tilt3} in the half-filled first excited 
($n=1$) LL.
The incompressibility at $\nu=2+{1\over2}={5\over2}$ could not be 
explained within Haldane's hierarchy (or Jain's non-interacting CF) 
picture and it was immediately obvious that it implied a different
type of correlations.
Since even-denominator Laughlin states occur for bosons, electron 
pairing was suggested by Halperin,\cite{halperin-pair} and various 
explicit paired-state trial wavefunctions have been constructed by 
a number of authors.\cite{haldane-pseudo,morf-pair,moore,greiter}
Although earlier theories\cite{haldane-pseudo,belkhir} suggested 
$s$-pairing (spin depolarization due to a small Zeeman energy; 
an idea later seemingly supported by experiments in tilted magnetic 
fields\cite{eisenstein-tilt1,eisenstein-tilt2,eisenstein-tilt3}), 
it is now established\cite{morf-52,rezayi-52} that the $\nu={5\over2}$ 
state is well described by a spin-polarized wavefunction introduced 
by Moore and Read (MR).\cite{moore}
Morf\cite{morf-52} and Rezayi and Haldane\cite{rezayi-52} compared 
the actual Coulomb eigenstates of up to 16 electrons with different 
trial wavefunctions, and found that the $\nu={5\over2}$ GS has large 
overlap with the (particle--hole symmetrized) MR\cite{moore} state, 
the phase transition between the ``CF behavior'' and pairing is driven 
by the strength of interaction at short range, and the actual Coulomb 
pseudopotential in the $n=1$ LL is close to the transition point.

While the non-Laughlin character of the $\nu={5\over2}$ state follows
from Haldane's ``odd-denominator'' rule, the type of correlations 
that cause incompressibility of other FQH states observed
\cite{willet,eisenstein-review,eisenstein-tilt1,eisenstein-tilt2,%
eisenstein-tilt3} in the $n=1$ LL have not yet been completely
understood.
The occurrence of the FQH effect at such prominent Laughlin--Jain 
fractions as $\nu=2+{1\over3}={7\over3}$, $2+{2\over3}={8\over3}$, 
or $2+{1\over5}={11\over5}$ might indicate that, although weakened 
because of reduction of Coulomb repulsion at short range, Laughlin 
correlations persist in the excited ($n=1$) LL.
The decrease of excitation gaps (e.g., the gap at $\nu={7\over3}$ 
being smaller than at $\nu={1\over3}$) could be interpreted as a 
direct measure of this weakening, and it might seem natural that 
only the most prominent FQH states of the $n=0$ LL persist at $n=1$.
Consequently, one could try to model correlations in the excited LL's 
using some modified version of the hierarchy or CF picture.
For example, it has been proposed\cite{belkhir,park} that the CF's 
are formed in excited LL as well (i.e., the electrons bind vortices 
of the many-body wavefunction -- which is a definition of Laughlin 
correlations), although the effects of CF--CF interaction (pairing) 
are more important at $n=1$.
On the other hand, numerical calculations\cite{macdonald,wojs-parentage} 
seem to disagree with experiments by showing neither Laughlin 
correlations nor incompressibility at $\nu={7\over3}$.
For example, quite different energy spectra are obtained\cite{wojs-parentage} 
for $N\le11$ electrons at the same value of the LL degeneracy (flux) 
corresponding to the Laughlin $\nu={1\over3}$ filling of the $n=0$ 
and $n=1$ LL's.
In the $n=1$ LL, the Laughlin quasiparticles or the magneto-roton band 
do not occur, and the excitation gap oscillates as a function of $N$ 
and does not converge to a finite value for $N\rightarrow\infty$.

The occurrence of an incompressible GS at a specific filling factor
results from the type of correlations that generally occur in the 
low lying states near this filling.
Therefore, these correlations must be studied before the correct trial 
wavefunctions can be constructed (or, at least, before their success 
can be understood).
The correlations near the half-filling of the lowest and excited LL's 
are the main subject of this paper.
We assume complete spin-polarization of the partially filled LL and
perform the numerical calculations in Haldane's spherical geometry,
where each LL has the form of a $(2l+1)$-fold degenerate angular
momentum shell.
The correlations in a Hilbert space restricted to an isolated LL are 
best defined through the occupation numbers (fractional parentage
\cite{wojs-parentage,quinn,shalit,cowan}) ${\cal G}$ for different 
pair eigenstates labeled by the relative pair angular momentum 
${\cal R}$.
The ${\cal G}({\cal R})$ contains more information about the nature 
of a studied many-body state than its overlap with a trial wavefunction.
It is also easier to interpret than the real-space pair-correlation 
function $g(r)$.

We explain the effects of harmonic ($V_{\rm H}$) and anharmonic 
($V_{\rm AH}$) parts of the interaction pseudopotential $V=V_{\rm H}
+V_{\rm AH}$ on correlations, and formulate a simple theorem which 
links the ${\cal G}({\cal R})$ profile of low lying states with 
the sign of $V_{\rm AH}({\cal R})$.
The Laughlin correlations occur when $V_{\rm AH}({\cal R})>0$ and 
disappear when $V({\cal R})$ becomes harmonic at short range.
\cite{wojs-parentage,quinn}
This clarifies the physical meaning of the critical strength of the 
highest pseudopotential parameter (relative to the Coulomb value) 
at which the transition between the Laughlin and MR phases has been 
found.\cite{morf-52,rezayi-52} 

From the analysis of the energy spectra of $N\le16$ electrons at 
different values of $2l$ (LL degeneracy), we identify three series 
of non-degenerate ($L=0$) GS's which in the thermodynamic limit of 
$N\rightarrow\infty$ and $N/(2l+1)\rightarrow\nu$ converge to the 
incompressible states at $\nu={5\over2}$, ${7\over3}$, and ${8\over3}$.
As shown by Morf,\cite{morf-52} the finite-size MR $\nu={5\over2}$ 
states occur for even $N$ at $2l=2N+1$.
The $\nu={7\over3}$ state occurs at $2l=3N-7$, which is different 
than $2l=3N-3$ of the Laughlin $\nu={1\over3}$ state (the same is 
true for their particle--hole symmetric conjugates at $\nu={8\over3}$
and ${2\over3}$).

The analysis of the ${\cal G}({\cal R})$ curves obtained for different 
values of $N$ and $2l$ and different model pseudopotentials shows that 
the electron correlations near the half-filling of the $n=1$ LL depend 
critically on the harmonic behavior of $V({\cal R})$ at short range.
(at $\nu\le{9\over4}$ the CF picture with four attached fluxes works 
and for example the $\nu={11\over5}$ state has Laughlin correlations
\cite{wojs-parentage}).
Thus, the three incompressible states at $\nu={5\over2}$, ${7\over3}$, 
and ${8\over3}$ all have similar (not Laughlin electron--electron,
although maybe Laughlin pair--pair) correlations.
In all low lying states near the half-filling, electrons minimize the 
total parentage from two pair states of highest repulsion, ${\cal R}=1$ 
and 3, which results in ${\cal G}(1)\approx{\cal G}(3)$ and large
value of ${\cal G}(5)$.
Cusps in the dependence of ${\cal G}(1)+{\cal G}(3)$ and ${\cal G}(5)$ 
on $N$ and $2l$ coincide with occurrence of incompressible $\nu={5\over2}$, 
${7\over3}$, and ${8\over3}$ states (similar to cusps in ${\cal G}(1)$ 
and ${\cal G}(3)$ in the $n=0$ LL signalling the Laughlin--Jain states).
For the MR $\nu={5\over2}$ state, the number of ${\cal R}=1$ pairs is
roughly equal to the half of the electron number, ${1\over2}N$, which 
supports the conjecture of pairing.

In the second excited ($n=2$) LL, $V({\cal R})$ is sub-harmonic at 
short range and super-harmonic at long range, and the minimization
of energy requires avoidance of strongly repulsive pair states at 
the intermediate ${\cal R}=3$ and 5, that is having ${\cal G}(3)
\approx{\cal G}(5)<{\cal G}(1)\approx{\cal G}(7)$.
This is achieved by grouping of electrons into spatially separated 
$\nu=1$ droplets.
Our values of ${\cal G}(1)$ suggest that in a finite system each droplet 
consists of three electrons.
This precludes pairing in the $\nu={9\over2}$ state, but not formation
of larger droplets or the charge-density-wave stripe order
\cite{koulakov,rezayi-stripe} in an infinite system.

\section{Model}
%%%%%%%%%%%%%%%%%%%%%%%%%%%%%%%%%%%%%%%%%%%%%%%%%%%%%%%%%%%%%%%%%%%%%%%%
We consider a system of $N$ electrons confined on a Haldane sphere
\cite{haldane-hierarchy} of radius $R$.
The magnetic field $B$ normal to the surface is produced by a Dirac 
magnetic monopole placed at the origin.
The strength $2S$ of the monopole is defined in the units of flux 
quantum $\phi_0=hc/e$, so that $4\pi R^2B=2S\phi_0$ and the magnetic 
length is $\lambda=R/\sqrt{S}$.
The single-particle states (monopole harmonics)\cite{haldane-hierarchy,%
wu,fano} are the eigenstates of angular momentum $l\ge S$ and its 
projection $m$.
The single-particle energies fall into $(2l+1)$-fold degenerate angular 
momentum shells (LL's), and the $n$-th shell has $l=S+n$.

At large $B$, the electron--electron (Coulomb) interaction is weak 
compared to the cyclotron energy $\hbar\omega_c$, and the scattering
between different LL's can be neglected.
In the low lying many-electron states at a filling factor 
$\nu_{\rm tot}=2f+\nu$ (where $f$ is an integer and $\nu<1$), a number 
$f$ of lowest LL's (with $n=0$, 1, \dots, $f-1$) are completely filled.
For simplicity, in the following we will omit the subscript `tot' 
and, depending on the context, $\nu$ will denote either partial filling 
of the highest occupied LL or the total filling factor $\nu_{\rm tot}$. 

The Coulomb interaction within a partially filled LL (with $n=f$) is 
given by a pseudopotential\cite{haldane-pseudo,wojs-parentage,quinn} 
$V_{\rm C}^{(n)}({\cal R})$.
The pseudopotential $V({\cal R})$ is defined as the interaction energy 
$V$ of a pair of particles as a function of their relative angular 
momentum ${\cal R}$.
On a sphere, ${\cal R}=2l-L'$ where $L'=|{\bf l}_1+{\bf l}_2|$ is 
the total pair angular momentum. 
For identical (spin-polarized) fermions, ${\cal R}$ is an odd integer, 
and larger ${\cal R}$ means larger average separation.\cite{wojs-parentage}

The many-electron Hamiltonian can be written as
\begin{equation}
  H = \sum_{ijkl} 
      c_i^\dagger c_j^\dagger c_k c_l \left<ij|V|kl\right>
    + {\rm const}
\label{eq1}
\end{equation}
where $c_m^\dagger$ ($c_m$) creates (annihilates) an electron in state 
$\left|l=S+f,m\right>$ of the $n=f$ LL, the two body interaction matrix 
elements $\left<ij|V|kl\right>$ are related with $V({\cal R})$ 
through the Clebsch--Gordan coefficients.
The constant term includes the energy of the completely filled LL's with 
$n<f$, the cyclotron energy of the electrons in the $n=f$ LL, and their 
interaction with the underlying (rigid) completely filled LL's, and will 
be omitted.

Hamiltonian (\ref{eq1}) is diagonalized numerically in Haldane's 
spherical geometry, for a finite number $N$ of electrons at different 
values of $2l$, corresponding to ${1\over3}\le\nu{2\over3}$.
The result is the spectrum of energy $E$ as a function of total angular
momentum $L$.
The $L=0$ ground states (GS) separated from the rest of the spectrum 
by an excitation gap $\Delta$ represent the non-degenerate ($k=0$) GS's 
an a plane.
If a series of such GS's can be identified at increasing $N$ and 
$2l=\nu^{-1}N+{\rm const}$, and if the gap $\Delta$ does not collapse 
in the $N\rightarrow\infty$ limit, these GS's describe an incompressible 
state of an infinite 2DEG at a filling factor $2f+\nu$.

\section{Fractional parentage}
%%%%%%%%%%%%%%%%%%%%%%%%%%%%%%%%%%%%%%%%%%%%%%%%%%%%%%%%%%%%%%%%%%%%%%%%
The electric conductivity and other properties that involve electron 
scattering depend critically on the correlations in the partially 
filled LL, which in turn depend entirely on the form of interaction 
pseudopotential $V({\cal R})$.
The correlations are best described in terms of the coefficients of 
fractional (grand)parentage (CFGP)\cite{wojs-parentage,quinn,shalit,cowan} 
${\cal G}({\cal R})$.
The CFGP gives a fraction of electron pairs that are in the pair 
eigenstate of a given ${\cal R}$, and thus ${\cal G}({\cal R})$ can 
be regarded as a pair-correlation function.
The energy $E_{L\alpha}$ of a state $\left|L\alpha\right>$ can be 
conveniently expressed through CFGP's as 
\begin{equation}
   E_{L\alpha}={1\over2}N(N-1)
   \sum_{\cal R}{\cal G}_{L\alpha}({\cal R})V({\cal R}), 
\label{eq2}
\end{equation}
and the normalizaton condition is 
$\sum_{\cal R}{\cal G}_{L\alpha}({\cal R})=1$.
The CFGP's also satisfy another constraint,\cite{wojs-parentage,quinn} 
\begin{eqnarray}
   {1\over2}N(N-1)&&\sum_{\cal R}{\cal G}_{L\alpha}({\cal R})\,L'(L'+1)
\nonumber\\
   &&=L(L+1)+N(N-2)\,l(l+1),
\label{eq3}
\end{eqnarray}
where $L'=2l-{\cal R}$.

\section{Laughlin correlations}
%%%%%%%%%%%%%%%%%%%%%%%%%%%%%%%%%%%%%%%%%%%%%%%%%%%%%%%%%%%%%%%%%%%%%%%%
The pseudopotential $V_{\rm H}({\cal R})$ of the harmonic interaction 
$V_{\rm H}(r)\propto r^2$ within an isolated ($n$th) LL is linear in 
$L'(L'+1)$,\cite{wojs-parentage} and from Eqs.~(\ref{eq2}--\ref{eq3}) 
it follows that its energy spectrum is degenerate at each value of $L$.
In other words, the harmonic interaction (within an isolated LL) does 
not cause any correlations, which are hence entirely determined by
the anharmonic part $V_{\rm AH}({\cal R})$ of the total pseudopotential 
$V({\cal R})=V_{\rm H}({\cal R})+V_{\rm AH}({\cal R})$.
Moreover, at a filling factor $\nu\ge(2p+1)^{-1}$, most important is 
the behavior of $V({\cal R})$ at ${\cal R}\le2p+1$ (corresponding to 
the pair of ``nearest'' electrons in the Laughlin state) and at those
values where $V({\cal R})$ changes most quickly (i.e. where the 
``effective force'' $\sim{\rm d}V/{\rm d}\left<r\right>$ is the largest).
The occurrence of Laughlin correlations in the FQH systems and their 
insensitivity to the details of the pseudopotential result from the 
following:\cite{wojs-parentage,quinn}

\paragraph*{Theorem 1:}
If for any three pair states at ${\cal R}_1<{\cal R}_2<{\cal R}_3$ 
(i.e., at $L'_1>L'_2>L'_3$) the pseudopotential $V$ is super-harmonic 
(i.e., increases more quickly than linearly as a function of $L'(L'+1)$; 
i.e., $V_{\rm H}({\cal R}_1)=V_{\rm H}({\cal R}_3)=0$ and 
$V_{\rm H}({\cal R}_2)<0$), then the many-electron energy $E_L$ can be 
lowered (within a given Hilbert space $[N,2l,L]$) by transferring some 
of the parentage from ${\cal G}({\cal R}_1)$ and ${\cal G}({\cal R}_5)$ 
to ${\cal G}({\cal R}_3)$ without violating Eq.~(\ref{eq3}).
The same holds in the planar geometry, except that the harmonic 
pseudopotential on a plane is linear in ${\cal R}$.

As a result, if $V({\cal R})$ is a super-harmonic at small ${\cal R}$ 
(at short range), the lowest energy states at each $L$ have minimum 
possible parentage from the (most strongly repulsive) pair state 
at the smallest ${\cal R}$.
The complete avoidance of $p$ pair states at ${\cal R}<2p+1$ corresponds 
to a Jastrow $\prod_{i<j}(z_i-z_j)^{2p}$ in the many-electron wavefunction
and, in particular, the Laughlin incompressible $\nu=(2p+1)^{-1}$ GS
\cite{laughlin-fqhe} is the only state at a given $N$ and $2l$ for which 
${\cal G}(1)=0$.

\section{Pairing and Laughlin Paired States}
%%%%%%%%%%%%%%%%%%%%%%%%%%%%%%%%%%%%%%%%%%%%%%%%%%%%%%%%%%%%%%%%%%%%%%%%
If the pseudopotential is sub-harmonic at small ${\cal R}$ (i.e., at short 
range), for example $V_{\rm H}(1)=V_{\rm H}(5)=0$ and $V_{\rm H}(3)>0$, 
then it should be energetically favorable to minimize parentage from the 
${\cal R}=3$ state, even at the cost of a large value of ${\cal G}(1)$.
Although the resulting ${\cal R}=1$ pairs are not formed because of any 
electron--electron attraction, but rather because of repulsion from the 
surrounding 2DEG (and thus their stability depends on $\nu$), the 
many-electron correlations can be described in terms of electron pairing 
and the (possibly simpler) correlations between pairs.
On a sphere, each ${\cal R}=1$ pair is a boson with the total angular 
momentum of $l_2=2l-1$.
The two-boson pair states are labeled by the total angular momentum 
$L_2'=2l_2-{\cal R}_2$ where ${\cal R}_2$ is an even integer, and
the pair--pair interaction is defined by an effective pseudopotential 
$V_2({\cal R}_2)$.
The Pauli exclusion principle applied to individual electrons results
in a hard core at a number $p_2=2$ of lowest values of ${\cal R}_2$
(similar to that of charged excitons\cite{wojs-xminus}), so that 
${\cal R}_2\ge2p_2$ for all pairs.
Such hard core can be accounted for by a mean field (MF) composite boson 
(CB) transformation with $2p_2$ flux quanta attached to each boson.
The CB transformation gives an effective CB angular momentum $l_2^*
=l_2-p_2(N_2-1)$, where $N_2$ is the number of pairs.
In the CB picture, all many-boson $L$-multiplets can be obtained by 
addition of $N_2$ angular momenta $l_2^*$ of individual CB's (without
an additional hard core).
For example, the $\nu=1$ state of electrons corresponds to the condensate 
of CB's in their only available $l_2^*=0$ state.
If the pair--pair pseudopotential $V_2({\cal R}_2)$ is super-harmonic
(and $l_2^*>0$), an additional MF CB transformation attaching an even 
number of $2q_2$ fluxes to each pair can be applied to select the lowest 
energy band of paired states which avoid a number of $q_2$ lowest values 
of ${\cal R}_2$ beyond the hard core.
The electron and CB filling factors, in the $N\rightarrow\infty$ limit 
defined as $\nu=N/2l$ and $\nu_2^*=N_2/2l_2^*$, are related by 
\begin{equation}
   \nu^{-1}=(4\nu_2^*)^{-1}+1
\label{eq4}
\end{equation}
and, for example, the series of Laughlin correlated CB states at 
$\nu_2^*={1\over8}$, ${1\over6}$, ${1\over4}$, and ${1\over2}$ occur 
at the electron filling factors $\nu={1\over3}$, ${2\over5}$, ${1\over2}$, 
and ${2\over3}$, respectively.
It is quite remarkable that, coincidentally, some of the most prominent 
odd-denominator Laughlin--Jain fractions occur among these states along 
with the (even-denominator) half-filled state.

On a Haldane's sphere, Laughlin $\nu_2^*=(2q_2)^{-1}$ states of bosons 
have $2l_2^*=2q_2(N_2-1)$, and thus the Laughlin-correlated paired 
$\nu=2/(q_2+2)$ states occur at
\begin{equation}
   2l={q_2+2\over2}N-1-q_2.
\label{eq5}
\end{equation}
It is noteworthy that applying the particle--hole symmetry 
($N\leftrightarrow N_h$, where $N_h=2l+1-N$ is the number of holes in 
the isolated LL) to Eq.~(\ref{eq5}) generates a different series of 
states at 
\begin{equation}
   2l={q_2+2\over q_2}N+1.
\label{eq6}
\end{equation}
That is because (in a finite system on a sphere) Laughlin paired states of 
electrons at $\nu$ do not occur at the same values of $2l$ as the Laughlin 
paired states of holes at $1-\nu$ (a similar effect was discussed in 
Ref.~\onlinecite{marinescu}).

If only a fraction $2N_2/N<1$ of electrons formed pairs in a many-electron 
state, the correlations should be described in terms of $N_2$ pairs (bosons) 
and $N_1=N-2N_2$ excess electrons (fermions).
The pair states of one electron and one electron pair are labeled by
$L_{12}'=l_1+l_2-{\cal R}_{12}$ where $l_1\equiv l$ and ${\cal R}_{12}$ 
is any integer, and the electron--pair interaction is defined by 
$V_{12}({\cal R}_{12})$.
A multi-component MF composite particle (CP) transformation can be used to 
account for the electron--pair hard core which forbids ${\cal R}_{12}<2$.
In such transformation,\cite{wojs-xminus} each electron couples to 
$p_{12}=2$ flux quanta attached to each pair, and each pair sees equal 
number $p_{12}$ of fluxes attached to each electron (in addition to 
$2p_2$ fluxes that each pair sees on every other pair), giving CF and CB 
angular momenta $l_2^*=l_2-{1\over2}p_{12}N_1-p_2(N_2-1)$ and $l_1^*=l_1-
{1\over2}p_{12}N_2$.
It is easy to check that a full shell of $N=2l+1$ electrons (the $\nu=1$ 
state) can be viewed as the only available state of $N_2$ pairs and 
$N_1=2l+1-2N_2$ excess electrons, in which the pairs condense at 
$l_2^*=0$ and the electrons completely fill their CF shell of 
$2l_1^*=N_1-1$.

If both electron--pair and pair--pair repulsions are super-harmonic, 
additional CP transformations can be used to select low energy states 
in which an appropriate number of electron--electron, pair--pair, and 
electron--pair pair states at the smallest ${\cal R}_1$, ${\cal R}_2$, 
and ${\cal R}_{12}$, respectively, are avoided.
While the discussion of the multi-component electron--pair (boson--fermion) 
liquids with Laughlin correlations will be presented elsewhere,\cite{tbp} 
let us note that such a state might be a more appropriate description of 
the $\nu={7\over3}$ state than a fully paired $\nu_2^*={1\over8}$ state.

\begin{figure}[t]
\epsfxsize=3.40in
\epsffile{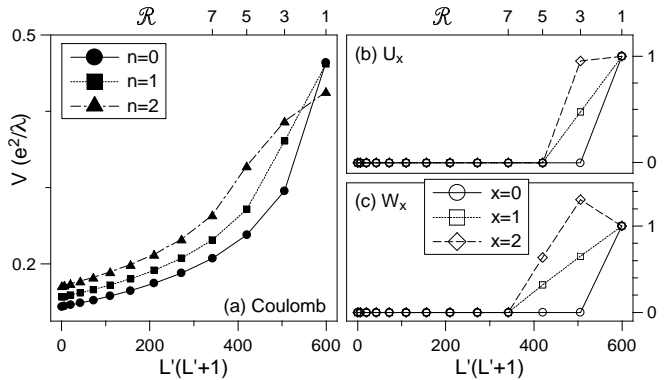}
\caption{
   The pseudopotentials (energy vs.\ squared pair angular momentum) 
   of the Coulomb interaction $V_{\rm C}^{(n)}$ in the lowest ($n=0$) 
   and two excited ($n=1$ and 2) Landau levels (a), and of the model 
   interactions $U_x$ (b) and $W_x$ (c), calculated for Haldane's 
   sphere with $2l=25$.
   $\lambda$ is the magnetic length.}
\label{fig1}
\end{figure}

The idea of a paired incompressible GS at $\nu={5\over2}$ (half-filled 
$n=1$ LL) has been suggested by a number of authors,\cite{haldane-pseudo,%
halperin-pair,morf-pair,moore,greiter} as the even-denominator fractions 
are characteristic of Laughlin-correlated systems of bosons.
However, as shown in Fig.~\ref{fig1}(a), the Coulomb pseudopotential 
$V_{\rm C}^{(1)}({\cal R})$ in the first excited LL is almost harmonic 
(linear in $L'(L'+1)$) rather than sub-harmonic between ${\cal R}=1$ 
and 5, and super-harmonic at larger ${\cal R}$.
Whether the above-sketched CP picture correctly describes correlations 
in the $\nu={5\over2}$ state depends on whether the harmonicity (or 
weak super-harmonicity) of $V_{\rm C}^{(1)}({\cal R})$ at ${\cal R}\le5$ 
is sufficient to cause pairing.
If only the pairs are formed, the pair--pair repulsion will certainly 
be super-harmonic (for the relevant ${\cal R}_2$) because the Coulomb 
repulsion in the $n=1$ LL is sub-harmonic only for small ${\cal R}$,
and not for electrons that belong to different pairs.

It is noteworthy that inclusion of the effects of the finite width of 
the quasi-2D electron layer even enhances the harmonicity of the Coulomb 
pseudopotential at short range.
This is because the pseudopotential of the 3D Coulomb interaction 
$V(r,z)\propto1/\sqrt{r^2+z^2}$ in a quasi-2D layer of width $w$ can 
be well approximated by that of an effective 2D potential $V(r)\propto
1/\sqrt{r^2+d^2}$ with $d=w/5$, and because $V(r)\approx(1-r^2/2d^2)/d$ 
at small $r$.
One can expect that other effects (such as due to the LL mixing) 
are too weak to produce large anharmonicity, and thus that the actual 
pseudopotential that occurs in the experimental systems is indeed
nearly harmonic at ${\cal R}\le5$.

\section{Numerical energy spectra for the Coulomb pseudopotential}
%%%%%%%%%%%%%%%%%%%%%%%%%%%%%%%%%%%%%%%%%%%%%%%%%%%%%%%%%%%%%%%%%%%%%%%%
If an incompressible GS occurs in an infinite system at a certain 
filling factor $\nu$, and if the correlations responsible for the 
incompressibility have a finite (short) range $\xi$, then the $L=0$ 
(non-degenerate) GS's are expected to occur in sufficiently 
large ($R>\xi$) finite (spherical) systems for a series of electron
numbers $N$ and LL degeneracies $2l+1$, such that $N/(2l+1)\rightarrow
\nu$ for $N\rightarrow\infty$.
In particular, for the $\nu={1\over2}$ filling (of the $n=1$ LL; 
relevant for the $\nu={5\over2}$ state) we expect such series at 
$N/(2l+1)\rightarrow{1\over2}$, for which $N_h/N\rightarrow1$.
The excitation gaps $\Delta$ above the $L=0$ GS's are generally 
expected to decrease as a function of $N$ (as the size quantization 
weakens) but it must converge to a finite value $\Delta_\infty>0$ in 
the $N\rightarrow\infty$ limit.

We have calculated the energy spectra of up to 16 electrons filling 
${1\over3}\le\nu\le{2\over3}$ of the lowest, first excited, and second
excited LL.
Due to the particle--hole symmetry in an isolated LL ($N\leftrightarrow
N_h$), only the systems with $N_h\ge N$ need be considered.
The dependence of the GS degeneracy and excitation gap $\Delta$ on $N$ 
and $2l$ (i.e., on $N$ and $\nu$) is different in different LL's.
As pointed out by Morf,\cite{morf-52} near the half-filling of the $n=1$ 
LL the non-degenerate ($L=0$) GS's with the largest excitation gaps occur 
in systems with the even values of $N$ and $|N-N_h|=2$. 
This corresponds to even $N$ and $2l=2N-3$, the values for the MR 
$\nu={5\over2}$ state, or its particle--hole conjugate at $2l=2N+1$.
Indeed, these numerical GS's were shown\cite{morf-52} to have large 
overlap with the spherical version of the exact MR trial wavefunction.
Note also that, as given by Eq.~(\ref{eq6}), the value $2l=2N-3$ 
describes the Laughlin $\nu_2^*={1\over4}$ state of ${\cal R}=1$ pairs.
The excitation gaps for $N=N_h+2=10$, 12, 14, and 16 electrons are 
$\Delta=0.0192$, 0.0258, 0.0220, and 0.0219~$e^2/\lambda$, respectively.
A similar series of non-degenerate ($L=0$) GS's with slightly smaller 
gaps occur for all even values of $N=N_h$ (i.e., at $2l=2N-1$), except 
for $N=10$.
Both these series correspond to the half-filled $n=1$ level (i.e. to 
$\nu={5\over2}$) in the $N\rightarrow\infty$ limit.
In the following, we assume that the series of $N$-electron GS's at 
$2l=2N+1$ in the $n=1$ LL describes the $\nu={5\over2}$ state of an
infinite (planar) system, and study correlations in these states.

We have also identified two other series of non-degenerate GS's with 
fairly large excitation gaps.
One series occurs at both odd and even values of $N$ and at $2l=3N-7$, 
and these GS's correspond to the $\nu={7\over3}$ filling in the 
$N\rightarrow\infty$ limit.
The gaps for $N=8$, 9, \dots, 12 electrons are $\Delta=0.0192$,
0.0295, 0.0217, 0.0140, and 0.0049~$e^2/\lambda$, respectively.
From the particle--hole symmetry, the other, conjugate ($\nu={8\over3}$) 
series occurs at even values of $N$ and at $2l={3\over2}N+2$.
Note that neither of these series occur at the values of $2l$ given 
by Eqs.~(\ref{eq5}) or (\ref{eq6}) corresponding to the Laughlin 
paired $\nu_2^*={1\over8}$ (for $\nu={7\over3}$) or  $\nu_2^*={1\over2}$ 
(for $\nu={8\over3}$) state.

\section{Numerical energy spectra for model pseudopotentials}
\begin{figure}[t]
\epsfxsize=3.34in
\epsffile{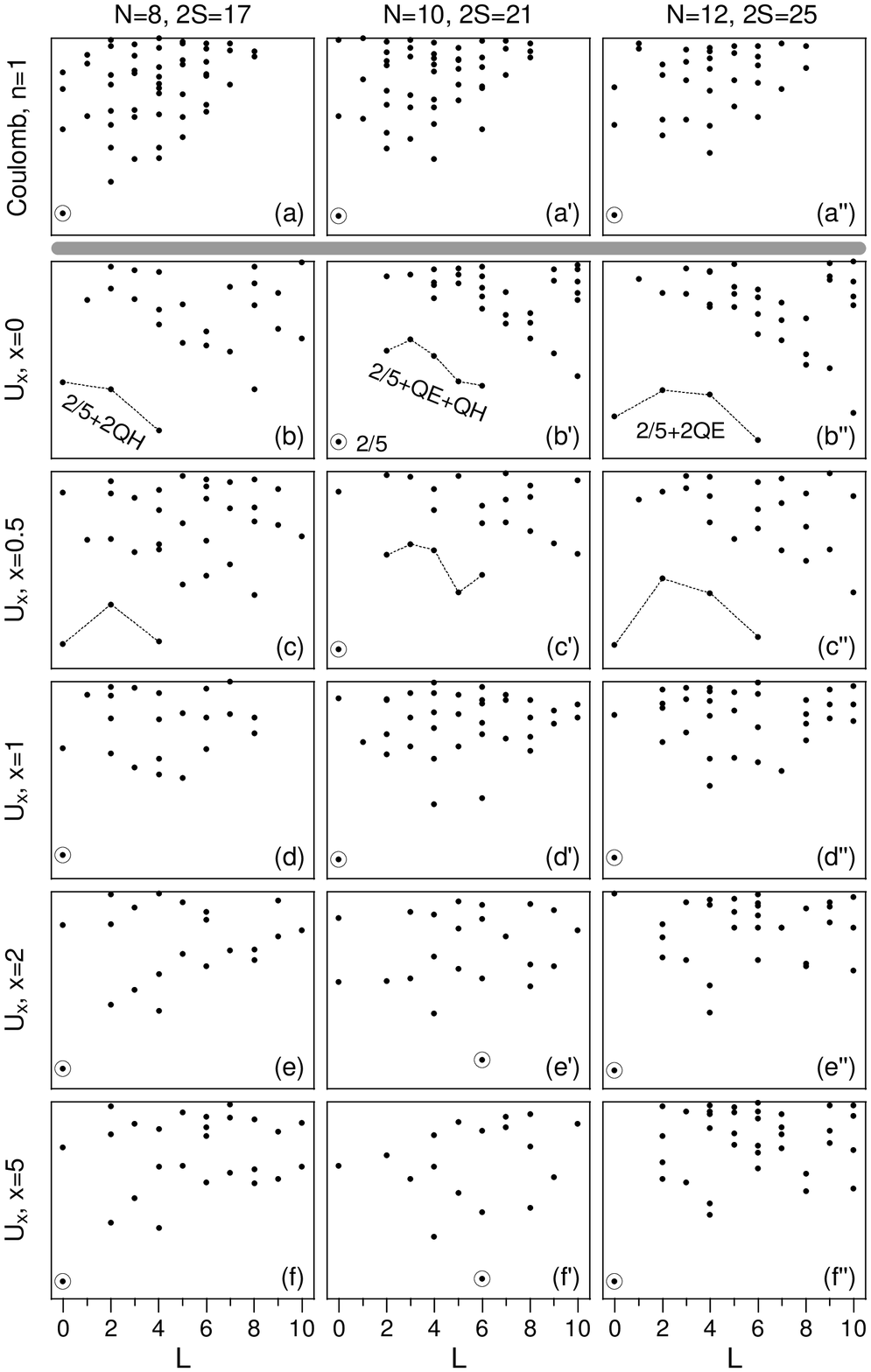}
\caption{
   The $N$-electron energy spectra (energy vs.\ angular momentum $L$) 
   on a Haldane's sphere:
   $N=8$ and $2l=17$ (a--f),
   $N=10$ and $2l=21$ (a$'$--f$'$), and
   $N=12$ and $2l=25$ (a$''$--f$''$),
   calculated for the Coulomb pseudopotential in the first excited
   Landau level $V_{\rm C}^{(1)}$ (a--a$''$), and for model interaction 
   $U_x$ with $x$ between 0 (b--b$''$) and 5 (f--f$''$).
   Circles and lines mark the lowest energy states.
   The Moore--Read $\nu={5\over2}$ state is the ground state in each 
   Coulomb spectrum.}
\label{fig2}
\end{figure}

%%%%%%%%%%%%%%%%%%%%%%%%%%%%%%%%%%%%%%%%%%%%%%%%%%%%%%%%%%%%%%%%%%%%%%%%
The pseudopotential of the Coulomb ($\propto r^{-1}$) interaction is 
different in different LL's.
For $n=0$ it is super-harmonic in the entire range of ${\cal R}$, while 
for $n=1$ it is super-harmonic at ${\cal R}\ge5$ but harmonic between 
${\cal R}=1$ and 5.
To study the transition of the electron system at $\nu\ge{1\over3}$ from 
the Laughlin- to MR-correlated phase we use a model pseudopotential 
$U_x({\cal R})$ shown in Fig.~\ref{fig1}(b), for which $U_x(1)=1$, 
$U_x({\cal R}\ge5)=0$, and $U_x(3)=x\cdot V_{\rm H}(3)$, where 
$V_{\rm H}(3)$ is the ``harmonic'' value defined so that $U_1$ is 
linear in $L'(L'+1)$ for ${\cal R}$ between 1 and 5.
The $U_x({\cal R})$ is intended to model the anharmonic part of a
repulsive (Coulomb) pseudopotential (at short range).
The omitted harmonic part does not affect many-electron wavefunctions 
and only results in a shift of the entire energy spectrum by a constant 
$\propto L(L+1)$.
The variation of $x$ in $U_x({\cal R})$ from $x=0$ through $x=1$ 
up to $x>1$ (super-harmonic, harmonic, and sub-harmonic at small 
${\cal R}$, respectively) allows calculation of wavefunctions and 
energy spectra of systems whose low energy states have well known 
correlations (Laughlin correlations at $x=0$ and pairing or grouping
into larger clusters at $x\gg1$), and their comparison with those of 
Coulomb pseudopotentials for different $n$. 
The comparison of the $n=1$ Coulomb energy spectra with the spectra
of $U_x$ with $x=0$, ${1\over2}$, 1, 2, and 5 is shown in 
Fig.~\ref{fig2} for the systems of $N=8$ (a--f), 10 (a$'$--f$'$), and 
12 electrons (a$''$--f$''$) at $2l=2N+1$, in which the MR GS occurs 
in the $n=1$ LL.
The energy scale is not shown on the vertical axes because the graphs 
are intended to show the structure of low energy spectra rather than 
the values of energy (the values obtained for the model pseudopotentials 
scale with $U_x(1)$, which we arbitrarily set equal to unity, and should 
include additional energy due to the neglected harmonic part of the 
pseudopotential).

In the spectra for $x<1$ (b--b$''$ and c--c$''$) the low lying states
have Laughlin correlations and can be understood within the CF (or 
Haldane's hierarchy) picture.
For the three systems used in our example, the lowest states are Jain 
$\nu={2\over5}$ GS at $L=0$ and the band of excited states at $2\le L\le6$ 
containing a quasielectron--quasihole (QE--QH) pair (b$'$--c$'$), and the 
states containing a pair of QH's (b--c) or QE's (b$''$--c$''$) in the 
$\nu={2\over5}$ state.

While it is well known that the energy spectra for $x<1$ are similar 
to the Coulomb spectra in the lowest LL, they are clearly different 
from those in the first excited LL.
As expected from the behavior of $V_{\rm C}^{(1)}({\cal R})$, the best 
approximation to the $n=1$ Coulomb spectra is obtained for $U_x$ with 
$x\approx1$.
Regardless of the value of GS angular momentum in the $x=0$ spectra, 
the $L=0$ GS's occur in all the three systems at $x=1$.
At $x\gg1$, when $U_x({\cal R})$ becomes strongly sub-harmonic between 
${\cal R}=1$ and 5, the $L=0$ GS persists in some systems (f and f$''$) 
but not in others (f$'$).

\begin{figure}[t]
\epsfxsize=3.34in
\epsffile{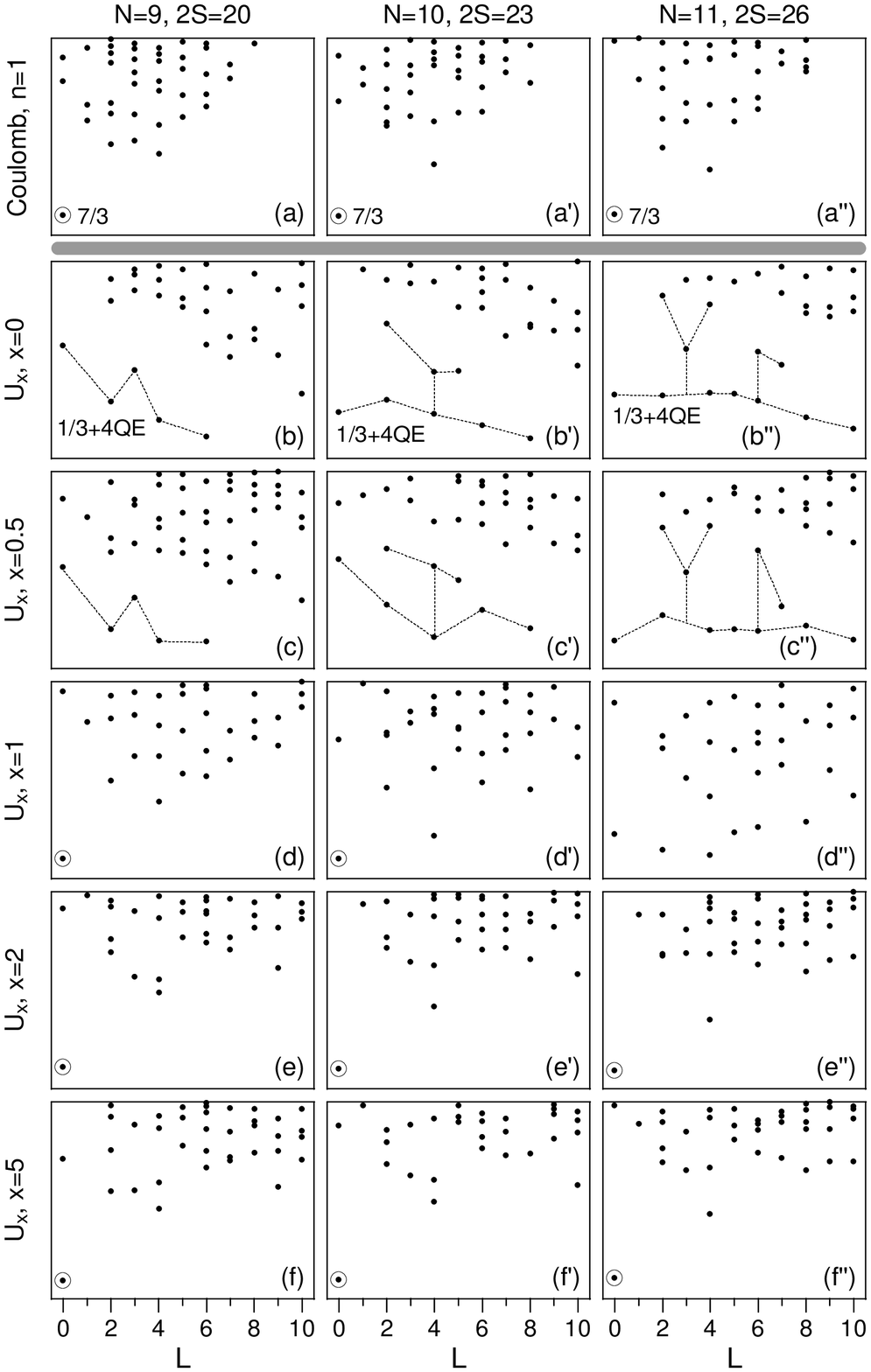}
\caption{
   The $N$-electron energy spectra (energy vs.\ angular momentum $L$) 
   on a Haldane's sphere:
   $N=9$ and $2l=20$ (a--f),
   $N=10$ and $2l=23$ (a$'$--f$'$), and
   $N=11$ and $2l=26$ (a$''$--f$''$),
   calculated for the Coulomb pseudopotential in the first excited
   Landau level $V_{\rm C}^{(1)}$ (a--a$''$), and for model interaction 
   $U_x$ with $x$ between 0 (b--b$''$) and 5 (f--f$''$).
   Circles and lines mark the lowest energy states.
   The incompressible $\nu={7\over3}$ state is the ground state in 
   each Coulomb spectrum.}
\label{fig3}
\end{figure}

Similar plots for the $\nu={7\over3}$ spectra of $N=9$, 10, 
and 11 electrons at $2l=3N-7$ are shown in Fig.~\ref{fig3}.
For each $N$, the low lying states of super-harmonic pseudopotentials $U_0$ 
(b--b$''$) and $U_{0.5}$ (c--c$''$) contain four QE's in the Laughlin 
$\nu={1\over3}$ state, while the Coulomb spectra in the $n=1$ LL (a--a$''$)
all have a $L=0$ ground state with a significant excitation gap, and all
resemble the spectra of harmonic and sub-harmonic pseudopotentials $U_1$
(d--d$''$), $U_2$ (e--e$''$), and $U_5$ (f--f$''$).

\section{Correlations in low lying states}
\begin{figure}[t]
\epsfxsize=3.40in
\epsffile{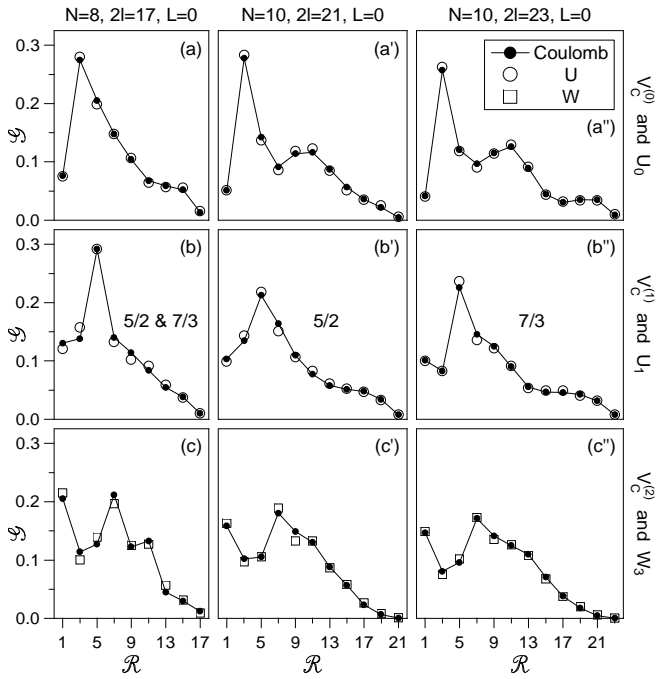}
\caption{
   The pair-correlation functions (coefficient of fractional 
   parentage ${\cal G}$ vs.\ relative pair angular momentum
   ${\cal R}$) in the lowest energy $L=0$ state of $N$ electrons 
   on a Haldane's sphere:
   $N=8$ and $2l=17$ (a--c),
   $N=10$ and $2l=21$ (a$'$--c$'$), and
   $N=10$ and $2l=23$ (a$''$--c$''$),
   calculated for the Coulomb pseudopotential in the lowest (a--a$''$),
   first excited (b--b$''$), and second excited (c--c$''$) Landau level,
   and for the appropriate model interaction $U_x$ or $W_x$.}
\label{fig4}
\end{figure}

%%%%%%%%%%%%%%%%%%%%%%%%%%%%%%%%%%%%%%%%%%%%%%%%%%%%%%%%%%%%%%%%%%%%%%%%
To find out if the correlations at $\nu={5\over2}$ or ${7\over3}$ can 
be understood in terms of electron pairing, we have analyzed the CFGP's 
of low lying states near the half filling.
In Fig.~\ref{fig4} we show some examples of the full ${\cal G}({\cal R})$ 
profiles (pair-correlation functions) calculated for the lowest $L=0$ 
states of eight and ten electrons at $2l=2N+1$ ($\nu={5\over2}$) and 
$2l=3N-7$ ($\nu={7\over3}$).
The $N=8$ state at $2l=17$ (a--c) contains two QH's in the incompressible 
$\nu={2\over5}$ state for the Coulomb interaction in the lowest LL, and 
it becomes a MR GS with a large excitation gap in the first excited LL.
The $N=10$ state at $2l=21$ (a$'$--c$'$) is the Jain $\nu={2\over5}$ state 
in the $n=0$ LL, and the MR state for $n=1$.
Finally, the $N=10$ state at $2l=23$ (a$''$--c$''$) contains four QE's in 
the Laughlin $\nu={1\over3}$ state in the $n=0$ LL, and it is the 
$\nu={7\over3}$ state for $n=1$.

\begin{figure}[t]
\epsfxsize=3.40in
\epsffile{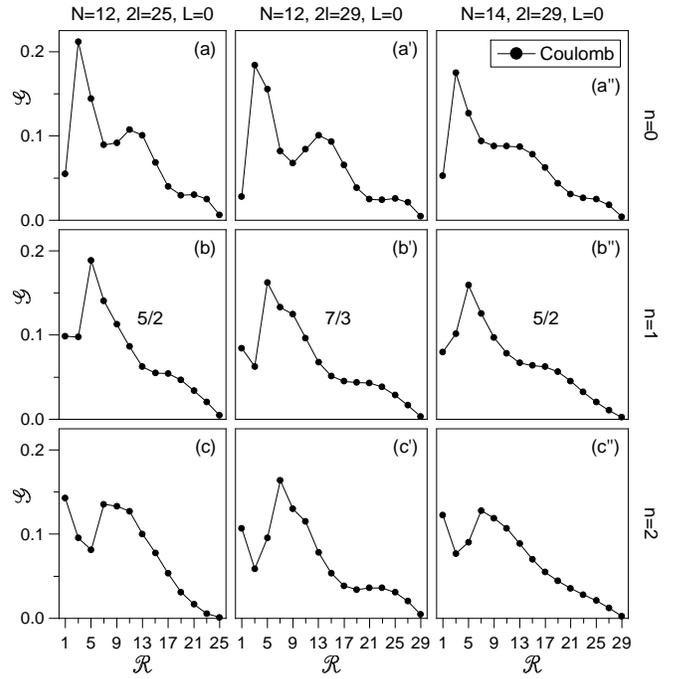}
\caption{
   The pair-correlation functions (coefficient of fractional 
   parentage ${\cal G}$ vs.\ relative pair angular momentum
   ${\cal R}$) in the lowest energy $L=0$ state of $N$ electrons 
   on a Haldane's sphere:
   $N=12$ and $2l=25$ (a--c),
   $N=12$ and $2l=29$ (a$'$--c$'$), and
   $N=14$ and $2l=29$ (a$''$--c$''$),
   calculated for the Coulomb pseudopotential in the lowest (a--a$''$),
   first excited (b--b$''$), and second excited (c--c$''$) Landau level.}
\label{fig5}
\end{figure}

It can be seen in Fig.~\ref{fig4}(a--a$''$) that for all three systems, 
the (Laughlin) correlations obtained for the $n=0$ Coulomb interaction 
are well reproduced by the model super-harmonic interaction $U_x$ with 
$x=0$ (the Laughlin correlations mean that the parentage ${\cal G}(1)$ 
from the ${\cal R}=1$ pair state is minimized).
From Fig.~\ref{fig4}(b--b$''$), the correlations in the $n=1$ LL are quite 
different, and they are better reproduced by the model interaction $U_x$ 
with $x=1$ (harmonic at short range).
Clearly, the Laughlin-like ``correlation hole'' at ${\cal R}=1$ 
characteristic of low lying states in the $n=0$ LL is absent for $n=1$.
Instead, the total parentage from the two states at ${\cal R}=1$ and 
3 is minimized, which results in the shift of the maximum of ${\cal G}
({\cal R})$ from ${\cal R}=3$ (as is for $n=1$) to ${\cal R}=5$.
Finally, the correlations for $n=2$ shown in Fig.~\ref{fig4}(c--c$''$) 
are not well reproduced by $U_x$ with any value of $x$.
A better approximation is obtained for a model pseudopotential $W_x({\cal 
R})$ shown in Fig.~\ref{fig1}(c), for which $W_x(1)=1$, $W_x({\cal R}\ge7)
=0$, $W_x(3)=x\cdot V_{\rm H}(3)$, and $W_x(5)=x\cdot V_{\rm H}(5)$, that 
is $W_x({\cal R})$ is harmonic between ${\cal R}=3$ and 7, and $x$ controls 
harmonicity between ${\cal R}=1$ and 5.
Similar plots for larger systems of $N=12$ and 14 electrons interacting 
through Coulomb pseudopotentials are shown in Fig.~\ref{fig5}.
In the $n=1$ LL, all three $L=0$ states in frames (b--b$''$) are the
incompressible ground states at $\nu={5\over2}$ or ${7\over3}$.

Let us note that a tendency of ${\cal G}$ to decrease with increasing 
${\cal R})$, observed most clearly at larger ${\cal R}$ (i.e., at 
separations beyond the correlation length), is characteristic of the 
closed (spherical) geometry.
For example, ${\cal G}$ decreases linearly as a function of ${\cal R})$ 
for the $\nu=1$ state).
However, the occurrence of minima and maxima in ${\cal G}({\cal R})$, i.e.
the differences between the values of ${\cal G}$ at neighboring values of
${\cal G}$, is independent of the geometry.

The above-described change of correlations when $n$ changes from 0 to 1
and 2 occurs for all low energy states (not only for the GS or the $L=0$ 
sector) and at any filling factor $\nu$ between about ${1\over3}$ and 
${2\over3}$.
Since the (Laughlin) correlation hole at small ${\cal R}$ results from 
the super-harmonicity of the pseudopotential at short range, it is not 
surprising that this hole changes from a single pair state at ${\cal R}=1$ 
(for $n=0$) to a couple of pair states at ${\cal R}=1$ and 3 (for $n=1$) 
or at ${\cal R}=3$ and 5 (for $n=2$), when the range of ${\cal R}$ in 
which the (Coulomb) pseudopotential is sub-harmonic changes with $n$.

\begin{figure}[t]
\epsfxsize=3.40in
\epsffile{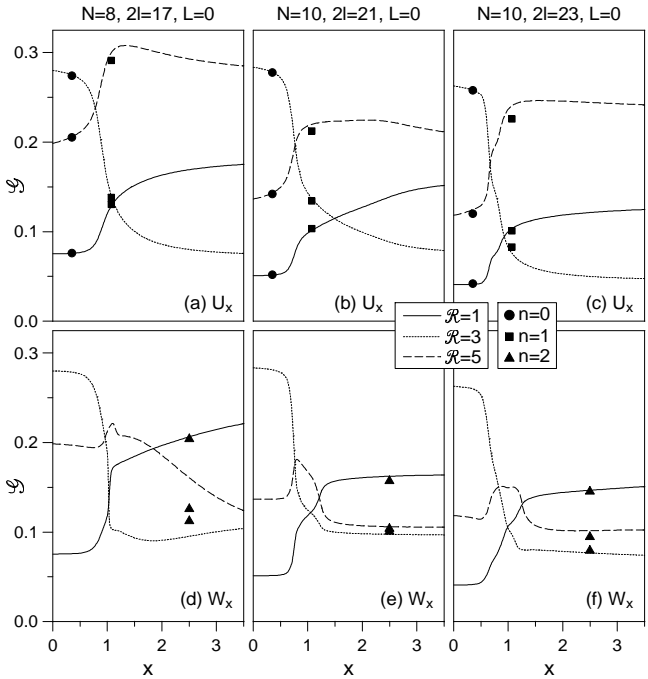}
\caption{
   The dependence of the coefficients of fractional parentage 
   ${\cal G}$ from pair states at the smallest values of relative 
   pair angular momentum, ${\cal R}=1$, 3, and 5, on the anharmonicity 
   parameter $x$ of the model pseudopotentials $U_x$ (abc) and $W_x$ 
   (def), calculated for the lowest $L=0$ state of $N$ electrons 
   on a Haldane's sphere:
   $N=8$ and $2l=17$ (ad),
   $N=10$ and $2l=21$ (be), and
   $N=10$ and $2l=23$ (cf).
   The values of ${\cal G}$ for the Coulomb pseudopotential in the lowest 
   and two excited Landau levels are marked with symbols.}
\label{fig6}
\end{figure}

The crossover between the Laughlin correlations and pairing is best 
observed in the dependence of the CFGP's at a few smallest values of 
${\cal R}$ on the anharmonicity parameter $x$ of the model interaction 
$U_x$.
In Fig.~\ref{fig6} we show the plots of ${\cal G}(1)$, ${\cal G}(3)$, 
and ${\cal G}(5)$ for the same lowest $L=0$ states as in Fig.~\ref{fig4},
that is states of eight electrons at $2l=17$ (a) and of ten electrons at 
$2l=21$ (b) and 23 (c), obtained for the $U_x$ interaction.
At $x<1$, when $U_x$ is super-harmonic in the entire range of ${\cal R}$,
the Laughlin correlations occur, meaning that ${\cal G}(1)$ is close to 
its minimum possible value.
As long as the interaction is super-harmonic (at short range), the values 
of CFGP's (and thus also the wavefunctions) weakly depend on the details 
of the pseudopotential (here, on $x$).
At $x>1$, correlations of a different type occur, which persist up to 
the $x\rightarrow\infty$ limit.
These correlations mean avoiding as much as possible the pair state at 
${\cal R}=3$ (i.e., the most super-harmonic part of $U_x$), which results 
in a large parentage from ${\cal R}=1$.
The abrupt crossover between the two types of correlations occurs near 
$x=1$, where ${\cal G}(1)$ quickly increases from its minimum value,
${\cal G}(3)$ drops to its minimum value, and a maximum occurs in 
${\cal G}(5)$.
At the crossing points in frames (ab), ${\cal G}(1)$ is close to the value 
$(N-1)^{-1}$ describing $N_2={1\over2}N$ pairs each with ${\cal R}=1$.
To obtain this value, which we will denote by ${\cal G}_{N_2\times2}(1)$, 
we use the fact that the contribution of each $\nu=1$ droplet of $N'$ 
electrons to the total number ${1\over2}N(N-1){\cal G}(1)$ of ${\cal R}=1$ 
pairs is ${1\over2}N'(N'-1){\cal G}_{1\times N'}(1)$, where the coefficient 
${\cal G}_{1\times N'}(1)$ describes an isolated droplet.

The CFGP's calculated for the Coulomb pseudopotentials with $n=0$ and 1 
are marked in Fig.~\ref{fig6} with full symbols.
The symbols are plotted at arbitrary values of $x$ to show that the 
correlations for $V_{\rm C}^{(0)}$ can be well reproduced by $U_x$ with 
a finite $x<1$, and that the correlations for $V_{\rm C}^{(1)}$ are well 
approximated by $U_x$ with $x\approx1$.

The most important conclusion from Fig.~\ref{fig6} is that the correlations 
in the partially filled (in particular, half-filled) LL are very sensitive
to the harmonicity of the pseudopotential at short range, and the largest 
(smallest) number of pairs occurs at those of small values of ${\cal R}$, 
at which $V({\cal R})$ is sub- (super-) harmonic.
The Coulomb pseudopotential $V_{\rm C}^{(1)}$ in the $n=1$ LL is nearly 
harmonic between ${\cal R}=1$ and 5, and thus the correlations it causes 
correspond to the crossover point between the sub- and super-harmonic 
regimes.
The number of ${\cal R}=1$ pairs in the (MR) GS at $\nu={5\over2}$ is 
almost equal to half the number of electrons, ${1\over2}N$.
This is consistent with the notion of the paired character of the (MR) 
ground state, and supports its interpretation at the Laughlin paired
$\nu_2^*={1\over4}$ state.
The $\nu={7\over3}$ GS shown in Fig.~\ref{fig6} does not occur at the 
value of $2l$ given by Eq.~(\ref{eq5}) or (\ref{eq6}).
Also, the value of ${\cal G}(1)$ in this state seems smaller than 
${\cal G}_{N_2\times2}(1)$.
This precludes a description of this state as involving Laughlin 
correlations among ${1\over2}N$ electron pairs each with ${\cal R}=1$.

The correlations induced by $V_{\rm C}^{(2)}$ are different from those 
in the $n=0$ or $n=1$ LL and cannot be modeled by $U_x$.
The reason is that $V_{\rm C}^{(2)}$ is not super-harmonic up to 
${\cal R}=7$.
A better approximation is obtained using model pseudopotential 
$W_x({\cal R})$.
The plots of ${\cal G}(1)$, ${\cal G}(3)$, and ${\cal G}(5)$ for 
the $W_x$ interaction in Fig.~\ref{fig6}(def) show a similar break-up
of Laughlin correlations at $x\approx1$ as those for $U_x$.
It is clear that the correlations in the $n=2$ LL can be modeled
by $W_x$ with an appropriate $x>1$, and also that the effective value 
of $x$ (i.e., the correlations) depends on $\nu$.
It can be expected that the tendency to occupy the ${\cal R}=1$ state 
and to avoid the ${\cal R}=3$ and 5 states will cause grouping of 
electrons into larger $\nu=1$ droplets.
Indeed, the values of ${\cal G}(1)$ for the Coulomb states in 
Fig.~\ref{fig6}(def) are much larger than ${\cal G}_{N_2\times2}(1)$.

\begin{figure}[t]
\epsfxsize=3.40in
\epsffile{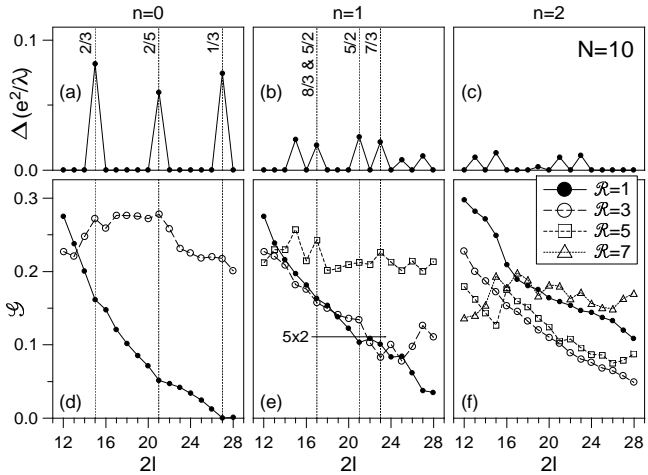}
\caption{
   The dependence of the excitation gap (abc) and the coefficients of 
   fractional parentage ${\cal G}$ from pair states at the smallest 
   values of the relative pair angular momentum, ${\cal R}=1$, 3, 5, and 
   7 (def) on $2l$, calculated for the ground states of $N=10$ electrons 
   on a Haldane's sphere, in the lowest (ad), first excited (be), and 
   second excited (cf) Landau levels.
   For degenerate ground states ($L\ne0$) the gap is set to zero.}
\label{fig7}
\end{figure}

\begin{figure}[t]
\epsfxsize=3.40in
\epsffile{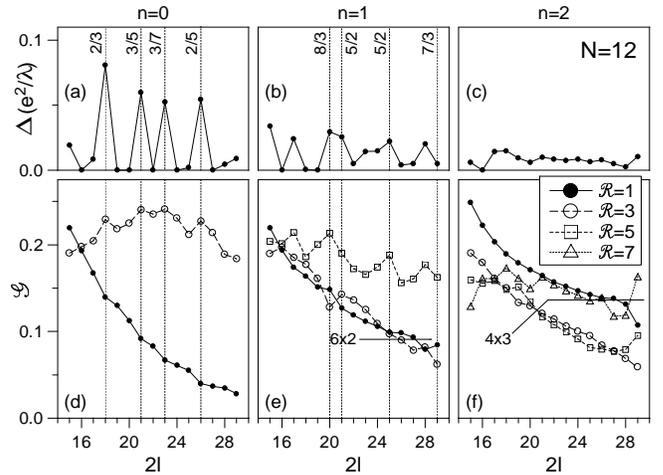}
\caption{
   The dependence of the excitation gap (abc) and the coefficients of 
   fractional parentage ${\cal G}$ from pair states at the smallest 
   values of the relative pair angular momentum, ${\cal R}=1$, 3, 5, and 
   7 (def) on $2l$, calculated for the ground states of $N=12$ electrons 
   on a Haldane's sphere, in the lowest (ad), first excited (be), and 
   second excited (cf) Landau levels.
   For degenerate ground states ($L\ne0$) the gap is set to zero.}
\label{fig8}
\end{figure}

More insight into the nature of correlations in different LL's can be 
obtained from Figs.~\ref{fig7} and \ref{fig8}, in which we plot the 
dependences of the excitation gap $\Delta$ and parentage coefficients 
${\cal G}({\cal R})$ for a few smallest values of ${\cal R}$ on the 
value of $2l$ (i.e., on $\nu$).
The gaps $\Delta$ are taken from the $L=0$ GS's, and we set $\Delta=0$ 
when the GS has $L\ne0$.
The CFGP's are calculated for the absolute GS's of $N$ electrons at 
given $2l$ (not the lowest energy $L=0$ state).

The comparison of curves for $N=10$ and 12 confirms that to minimize 
total interaction energy at any $\nu$, electrons interacting through 
a pseudopotential $V({\cal R})$ avoid as much as possible the total 
parentage from pairs states corresponding to $V_{\rm AH}({\cal R})>0$.
Because of relation (\ref{eq3}), minimization of parentage from those 
most strongly repulsive pair states implies large parentage from less 
strongly repulsive pair states at the neighboring values of ${\cal R}$.
Thus, for $n=0$ the occurrence of incompressible Laughlin--Jain states
with large $\Delta$ coincides with downward peaks in ${\cal G}(1)$ and
upward peaks in ${\cal G}(3)$.
For $n=1$, where ${\cal G}(1)+{\cal G}(3)$ is minimized, large $\Delta$ 
coincides with upward peaks in ${\cal G}(5)$.
Finally, for $n=2$ the occurrence of gaps seems to be connected with the 
behavior of ${\cal G}(7)$.

Note that in the $n=1$ LL, the gap $\Delta=0.0049$~$e^2/\lambda$ in the 
$N=12$ electron system at $2l=29$ is smaller than the gaps for $N\le11$ 
at the same filling factor (given by $2l=3N-7$) and than the gap for 
$N=12$ at a neighboring $2l=28$.
The diminishing of $\Delta$ as a function of $N$ in the $2l=3N-7$ series
of GS's indicates that this series might not describe the observed 
incompressible $\nu={7\over3}$ state in the $N\rightarrow\infty$ limit.
In any case, it remains true that the occurrence of a finite-size $L=0$ 
GS with a large gap ($\Delta=0.0201$~$e^2/\lambda$) at $N=12$ and $2l=28$ 
coincides with an upward cusp in ${\cal G}(5)$.

The occurrence of similar maxima in ${\cal G}(5)$ at $\nu={5\over2}$,
${7\over3}$, and ${8\over3}$ (or, more exactly, at the values of $N$ and 
$2l$ at which non-degenerate GS's with large gaps occur) for $n=1$ 
indicates common correlations in these three states, different from 
those in other LL's.
We have marked the values of ${\cal G}(1)$ corresponding to grouping of 
$N$ electrons into ${1\over2}N$ pairs, ${\cal G}_{N_2\times2}(1)=(N-1)^{-1}$.
Clearly, the average number of ${\cal R}=1$ pairs decreases with 
increasing $2l$ which seems to disagree with the prediction of Laughlin 
paired $\nu_2^*=(2q_2)^{-1}$ states for all values of $q_2$ between 1 
and 4 (for Laughlin paired states one should expect 
${\cal G}(1)\approx(N-1)^{-1}$ independently of $2l$).
However, the number of ${\cal R}=1$ pairs is roughly equal to ${1\over2}N$
for $2l$ corresponding to the MR state at $\nu={5\over2}$, which suggests 
the Laughlin paired $\nu_2^*={1\over4}$ state as an appropriate description 
at this particular filling.

The observation that ${\cal G}(1)$ in the $n=1$ LL decreases monotonically 
as a function of $2l$ and that ${\cal G}(1)\approx(N-1)^{-1}$ at 
$\nu={5\over2}$ suggests that all $N$ electrons form pairs at exactly 
$\nu={5\over2}$, but only a fraction of electrons pair up ($N_2<{1\over2}N$ 
and $N_1>0$) when $\nu<{5\over2}$, and some pairs are replaced by larger 
$\nu=1$ clusters (e.g., into three-electron droplets each with $l_3=3l-3$) 
when $\nu>{5\over2}$.
The break-up or clustering of pairs can be understood from the behavior 
of the effective pseudopotentials describing interaction between electrons, 
pairs, and larger droplets, and will be discussed in a subsequent 
publication.\cite{tbp}

In the $n=2$ LL, the average number of ${\cal R}=1$ pairs is larger
than ${1\over2}N$, indicating formation of larger $\nu=1$ droplets 
(stripes\cite{koulakov,rezayi-stripe}) separated from one another.
As marked in Fig.~\ref{fig8}(f), in the (fairly small) $N=12$ electron 
system, ${\cal G}(1)\approx{\cal G}_{4\times3}(1)={3\over2}(N-1)^{-1}$ 
near the half-filling, which corresponds to four three-electron droplets.

\section{Conclusion}
%%%%%%%%%%%%%%%%%%%%%%%%%%%%%%%%%%%%%%%%%%%%%%%%%%%%%%%%%%%%%%%%%%%%%%%%
Using exact numerical diagonalization in Haldane's spherical geometry 
we have studied electron correlations near the half-filling of the lowest
and excited LL's.
We have shown that the electrons interacting through a pseudopotential 
$V({\cal R})$ generally avoid pairs states corresponding to large and 
positive anharmonicity of $V({\cal R})$.
We have shown that as a result of different behavior of $V({\cal R})$
in different LL's, the correlations in the excited LL's are different 
than the Laughlin correlations in the lowest LL.
This confirms different origin of the incompressibility of the 
$\nu={1\over3}$ and ${7\over3}$ GS's.
In particular, correlations in the partially filled first excited 
($n=1$) LL depend critically on the harmonic behavior of the Coulomb 
pseudopotential at short range, and are destroyed when the pseudopotential 
becomes either strongly super-harmonic (as for $n=0$) or strongly 
sub-harmonic (as for $n=2$).
The Moore--Read incompressible state at $\nu={5\over2}$ occurs at the LL 
degeneracy (flux) given by $2l=2N-3$ (and $2l=2N+1$ for its particle--hole
conjugate).
This value of $2l$ and the calculated CFGP's for the low lying states 
indicate that the Moore--Read $\nu={5\over2}$ state can be understood 
as a Laughlin correlated $\nu_2^*={1\over4}$ bosonic state of electron 
pairs.
Although other filling factors at which incompressibility is observed
in the $n=1$ LL ($\nu={7\over3}$ and ${8\over3}$) also arise in the
sequence of Laughlin paired $\nu_2^*=(2q_2)^{-1}$ states, we find no 
evidence that these are the actual Coulomb GS's.
The two series of finite-size non-degenerate GS's that we find in 
our numerical calculations and that extrapolate to $\nu={7\over3}$ 
and ${8\over3}$ for $N\rightarrow\infty$ occur at $2l=3N-7$ and 
${3\over2}N+2$.
These values of $2l$ are different from both these of Laughlin--Jain 
GS's at $\nu={1\over3}$ and ${2\over3}$ in the $n=0$ LL, and those of 
the hypothetical Laughlin paired states at $\nu_2^*={1\over8}$ and 
${1\over2}$.

\section{Acknowledgment}
%%%%%%%%%%%%%%%%%%%%%%%%%%%%%%%%%%%%%%%%%%%%%%%%%%%%%%%%%%%%%%%%%%%%%%%%
The author gratefully acknowledges helpful discussions with John J. Quinn 
(Univ.\ Tennessee), partial support of Grant DE-FG02-97ER45657 from the 
Materials Science Program -- Basic Energy Sciences of the US Department 
of Energy, and of Grant 2P03B11118 from the Polish Sci.\ Comm., and thanks 
the Joint Institute for Computational Science at the Univ.\ Tennessee for 
providing access to the IBM SP2 supercomputer and for user support.

\end{document}